\documentstyle[aps,preprint,prl]{revtex}

\begin{document}

\title{Odd time magnetic correlations and chiral spin nematics
}
\author{
A.\ V.\ Balatsky\protect\cite{aa}
}
\address{
 Theoretical Division, Los Alamos National Laboratory, Los Alamos, NM
87545}
\author{Elihu Abrahams}
\address{Serin Physics Laboratory, Rutgers Universtiy, P.O. Box 849, NJ
08855}
\date{\today}

\maketitle

\begin{abstract}
The magnetic analog of odd-time superconducting order is introduced. It
is shown that the classification of possible odd-time magnetic states
admits the existence of a novel state - the {\it chiral spin nematic} -
which is characterized by an odd-in-time spin-spin correlation function.
The known example of the {\it chiral spin liquid} appears naturally in
this approach. The equal-time correlation function of these odd-in-time
states involves non-trivial three spin correlations.
\end{abstract}


The question of broken symmetry states characterized by some correlation
function which is {\em odd} in time, was revived recently in the context
of
superconductivity.
An odd-time, or equivalently, odd-frequency, superconducting ordering
was introduced in the pioneering work of Berezinskii \cite{Ber}, who
considered spin-triplet odd-time pairing as an alternative to the
conventional theory of $^3$He pairing. Interest in odd-time
ordering was renewed after the work of Balatsky and Abrahams \cite{BA},
in which the extension of Ref. [1] to odd-time pairing for spin-singlet
superconductors was given.

In this paper, in analogy to odd-time superconductivity, we introduce the
concept of magnetic states with odd-time
correlations. We shall discuss the symmetry constraints for such states
and
shall classify the different possibilities.

For odd-time superconductivity, the
possibility of odd-time states follows immediately
from the basic symmetry constraint \cite{BA}  $F({\bf k},t) = F(-{\bf k}, -t)$
for the general spin-singlet anomalous correlator
$F({\bf k}, t) = i(\sigma^y)_{\alpha \beta} \langle~T_t c_{{\bf k}
\alpha}(t)
c_{-{\bf k} \beta}(0)\rangle$. This symmetry constraint is an immediate
consequence of the Fermi statistics of the $c_{{\bf k} \alpha}$
operators and of the singlet nature of the anomalous correlator. The
consideration for
spin-triplet pairing is
analogous and was done by Berezinskii \cite{Ber}. Then one finds
two classes of states:
1) even-time pairing (BCS) states and 2) odd-time
pairing states with $F({\bf k,} -t) = -F({\bf k}, t)$. The
orbital parities of these states are opposite [even(odd) and odd(even) for
singlet(triplet) respectively] as follows
from the symmetry equation.

The generalization of odd-time ordering to spin systems is
straightforward and requires the symmetry equation for the dynamic
correlation function of the spin density $S_i({\bf r},t)$ where $i= 1,2,3$
denotes cartesian components. The spin-spin correlation function is
$\Lambda_{ij}({\bf r},{\bf r'}|t) = \langle T_t ~
S_i({\bf r},t)S_j({\bf r'},0)\rangle $. The symmetry equation which
follows directly from the properties of the time-ordering operator in the
definition of $\Lambda_{ij}$ is
$\Lambda_{ij}({\bf r},{\bf r'}|t) =
\Lambda_{ji}({\bf r'},{\bf r}|-t)$. In matrix form, this is:
\begin{equation}
P T ~ {\tilde \Lambda}({\bf r},{\bf r'}|t) = \Lambda ({\bf r},{\bf r'}|t),
\label{basic}
\end{equation}
where $P$ is the spatial parity operator ${\bf r
- r'}\rightarrow {\bf r' - r}$, $T$ is the time inversion and ${\tilde
\Lambda}$ is the transposed matrix ${\tilde \Lambda}_{ij} =
\Lambda_{ji}$. This equation, valid for any rank spin $S$, allows the
existence of odd-time magnetic states. These states are characterized by a
nontrivial time  structure of the spin-spin correlation function with
$\Lambda_{ij}({\bf r},{\bf r'}|t)$ being an odd function of time.

The main purpose of this paper is to
classify
magnetic states with odd-time magnetic correlations.
We shall give the symmetry analysis for possible odd-time magnets. It follows
closely that for even-time magnets which have
nontrivial spin correlation functions.
All the known chiral states are recovered  in this
approach: The chiral spin liquid (CSL) and the ferromagnet (a rather dull
example of a chiral state) are among them. A new magnetic state is found
as well: It is a state which is the odd-time analog of the spin nematic. The
spin nematic was first considered by Andreev and Grishchuk \cite{AG}. The new
odd-time state is also characterized by nematic ordering in spin space
\cite{AG},  but with broken $T$  and $P$. We call this state a ``chiral spin
nematic" (CSN). We shall discuss the  physical
properties of this state and we shall show that the low energy lagrangian is
identical to that for a conventional spin
nematic \cite{AG}.

We begin with the symmetry of
the magnetic correlator. We consider states in which
spontaneous breakdown of the $O(3)$ spin rotation group occurs {\em
without} the appearance of an average microscopic spin density so that
$\langle S_i({\bf r},
t)\rangle = 0$. This restriction leads us to consider a spin-spin
correlation
function $\Lambda_{ij}({\bf r},{\bf r'}|t)$, which is time-dependent and
in general is a function of two coordinates ${\bf r},{\bf r'}$. As
shown by Andreev and Grishchuk
\cite{AG}, spin-spin correlations without time-reversal violation can
still describe a nontrivial magnetic order, i.e. a spin nematic. In that
case,
$\Lambda_{ij}({\bf r},{\bf r'}|0)$ transforms as a tensor
representation of $O(3)$. This will also be the case for odd-time
correlations described by $\Lambda_{ij}$ for $t \neq 0$.

Following the analysis of the Andreev and Grishchuk \cite{AG}, we consider
the invariance group of the system. It is given by the product
$O(3)\times T\times G$, where $O(3)$ is the spin rotation group, $G$ is
the space group of the lattice and $T$ is time reversal,
equivalent to spin inversion. $G$ defines the symmetry of
spin scalars; it contains the crystal point group and the translations of
the lattice.

$\Lambda_{ij}$ is a $3\times3$ matrix  containing 9 elements. It can be
decomposed as a sum of terms which transform as particular representations
of $O(3)$. From the decomposition 9=1+3+5 we find that $\l = 0,1,2$ can
be present. Therefore we write:
\begin{eqnarray}
\Lambda_{ij}({\bf r},{\bf r'}|t) = A({\bf r},{\bf r'}|t)\delta_{ij} +
\epsilon_{ijk}B_k({\bf r},{\bf r'}|t) + Q_{ij}({\bf r},{\bf r'}|t).
\label{decomposition}
\end{eqnarray}
where $A({\bf r},{\bf r'}|t)$ is  a scalar ($\l =0$)  and
is the only term which contributes to the trace of $\Lambda_{ij}({\bf
r},{\bf r'}|t)$. $B_k({\bf r},{\bf r'}|t)$ is a vector
($\l=1$) and  $Q_{ij}({\bf r},{\bf r'}|t)$ is the tensor part
($\l=2$). The latter is symmetric and traceless.
The quantities
$A$, $B_k$ and $Q_{ij}$ describe a spin liquid with no ordering,
magnetic ordering with a pseudovector order parameter and spin
nematic ordering, respectively. This is the case whatever
the parity and time reversal properties of these objects.

For the case of the even-time states, the order parameter can be
found from the equal-time correlator $\Lambda_{ij}({\bf r},{\bf r'}|t=0)$.
Then we recover the
 classification of Ref. \cite{AG}. In the even-time case,
$\Lambda_{ij}({\bf r},{\bf r'}|t)$ in  Eq.\ (\ref{decomposition}) is even
under time reversal. Then
$B_k({\bf r},{\bf r'}|t)$ and $ Q_{ij}({\bf r},{\bf r'}|t)$ describe  two
different nematic states, the so-called ${\bf p}$ and ${\bf n}$ nematics
respectively \cite{AG}. The $A$-term describes a spin liquid with no spin
ordering. Complete details for the even-time case may be found in Ref.\
\cite{AG}.

We now turn to odd-time correlations. In the case of
odd-time superconductivity, it is well-established
\cite{G4,Coleman,EK,BB} that the equal-time correlators which describe the
condensate contain composite operators which involve a conventional Cooper
pair bound with, for example, a spin-density operator. In the magnetic case,
we shall implement exactly the same approach and  the odd-time {\em
magnetic} correlators will describe the condensate of the higher order spin
operators. For any $\Lambda_{ij}({\bf r},{\bf r'}|t)$ which is odd in time,
the time derivative will be nonzero (we assume analyticity of
$\Lambda_{ij}$ at small
$t$) at $t=0$ and represents a thermodynamic order
parameter for a composite condensate:
\begin{equation}
\partial_{t}\Lambda_{ij}({\bf r},{\bf r'}|t)_{t=0} =
\langle T_t[\partial_{t}S_i({\bf r},t)] S_j({\bf
r'},0)\rangle_{t=0}
\end{equation}
The equation of motion for $S_i$ can be used to find the general result
\begin{equation}
\partial_t S_i({\bf r},t) = i [{\cal H},S_i({\bf r},t)] =
\epsilon_{ilm}S_l({\bf r},t) M_m({\bf r},t),
\end{equation}
where $M_m({\bf r},t)$ is the ``molecular field" for the hamiltonian ${\cal
H}$. If the hamiltonian is bilinear in spin
operators, the general form of the molecular field will be
\begin{equation}
M_m({\bf r}) = \int d{\bf r}' K_{mn}({\bf r},{\bf r}')  S_n({\bf r}').
\end{equation}
The coupling $K_{mn}({\bf r},{\bf r}')$ is an explicit function of the {\em
two}  co-ordinates ${\bf r}, {\bf r}'$.
Finally, we obtain
\begin{equation}
\partial_{t}\Lambda_{ij}({\bf r},{\bf r'}|t)_{t=0} = \epsilon_{ilm}\int
d{\bf r''} K_{mn}({\bf r},{\bf r''}) \langle  S_l({\bf r}) S_n({\bf r''})
S_j({\bf r'}) \rangle.
\label{deriv}
\end{equation}

If the hamiltonian has the
form
\begin{equation}
{\cal H} = - \sum_{mn} \int d{\bf r}d{\bf r}'S_m({\bf r})L_{mn}({\bf
r},{\bf r}') S_n({\bf r}'),
\label{example}
\end{equation}
then the molecular field kernel is given by
\begin{equation}
K_{mn}({\bf r},{\bf r}') = L_{mn}({\bf r},{\bf r}') + L_{nm}({\bf r}',{\bf
r}) = 2 L_{mn}({\bf r},{\bf r}').
\end{equation}
We see that any pure pairwise exchange terms of the hamiltonian will not
contribute to the time derivative of odd-time correlators, since the
corresponding kernel
$K_{mn}({\bf r},{\bf r''})$ for pairwise exchange is symmetric in ${\bf r},
{\bf r''}$, while
the correlator on the r.h.s of Eq.\ (\ref{deriv})
is antisymmetric. Therefore for a non-zero result, $K({\bf  r},{\bf r''})$ must
contain a
spatially-odd component. We shall give an explicit example of
a hamiltonian which produces these equations of motion for the specific case of
a
CSL. In general, complicated interactions are required for chiral magnetic
order.

We now discuss the specific phases which can be found in odd-time
magnets.
The possible terms in the decomposition Eq.\ (\ref{decomposition}) will be
considered in turn.

1). {\em  Chiral Spin Liquid}. Consider a spin-spin correlator
$\Lambda_{ij}$ containing only the $A$-term
 in the r.h.s. of Eq.\ (\ref{decomposition}). The spin scalar term $A({\bf
r},{\bf r'}|t)$ is odd in time and, therefore is odd under spatial parity
$P$, as can be easily seen from the symmetry equation (\ref{basic}) for any
cartesian-symmetric odd-time $\Lambda_{ij}$. Taking the time derivative as in
Eqs.\ (3-6), we find
\begin{equation}
\partial_{t}A({\bf r},{\bf r'}|0) = \int d{\bf r''} K({\bf r},{\bf r''})
\langle {\bf S}({\bf r})\times {\bf S}({\bf r''}) \cdot{\bf S}({\bf
r'})\rangle,
\label{csleq}
\end{equation}
where, for the assumed CSL state with no  anisotropy in spin space, we have
taken
$K_{mn}({\bf r},{\bf r''}) = \delta_{mn}K({\bf r},{\bf r''})$. Note that
$\partial_{t}A({\bf r},{\bf r'}|0)$ obtained in this way is
explicitly odd under $P$.

The necessary condition for the odd-time CSL phase  is the existence of
a non-zero  correlator $X_P({\bf r}_1{\bf r}_2{\bf r}_3) = \langle {\bf
S}({\bf  r}_1)\times {\bf S}({\bf r}_2) \cdot{\bf S}({\bf r}_3)\rangle$, which
when  defined on a plaquette is exactly the usual CSL order parameter
\cite{Wen}. We argue, therefore, that an alternative way to describe a CSL
phase is in terms of an odd-time spin singlet correlator $ A({\bf
r},{\bf r'}|t) = \langle T_t ~
S_i({\bf r},t)S_i({\bf r'},0)\rangle $. The well-known picture of the CSL
state as the state where all spins precess so that local handedness is
preserved is naturally supported in this  description.  In order for the
time derivative $\partial_{t}A(t=0)$ to exist, the coupling  $K({\bf
r},{\bf r'})$ should contain a part which is odd in ${\bf r}-{\bf r'}$. This
puts a severe constraint on the possible exchange models which can support the
CSL
ground state.  One of them is given by the hamiltonian \cite{comment1}
\begin{equation}
{\cal H} = -{\lambda\over{2}} \sum_{\langle ij\rangle} \left[{\bf S}_1\times
{\bf  S}_2 \cdot{\bf S}_3\right]_{P_i} \left[{\bf S}_4\times {\bf S}_5
\cdot{\bf S}_6\right]_{P_j},
\label{hamcsl}
\end{equation}
with $\lambda$-positive. The sum runs over nearest neighbor plaquettes
$P_i,P_j$ with spins $1,2,3$ and $4,5,6$ belonging to the plaquettes $P_i$ and
$P_j$ respectively \cite{Wen} and we consider here the triangular lattice.
This
hamiltonian  has a CSL ground  state with  ferromagnetic ordering of
chiralities  with nonzero  $X_{P}({\bf r}_1,{\bf r}_2,{\bf r}_3) $. It can be
shown that the equations  of motion obtained from this hamiltonian do indeed
lead to Eq.\  (\ref{csleq}) for $\partial_{t}A$, where the spin scalar kernel
$K({\bf r},{\bf  r'})$ is a certain combination of the pair spin  $\langle
{\bf S}({\bf r}_i) {\bf S}({\bf r}_j) \rangle$ correlation functions.

2). {\em Ferromagnet}. Consider the $B_k$ term in Eq.\
(\ref{decomposition}).
  $B_k$ transforms as a vector under the spin rotation group
 $O(3)$; it is even under spatial inversion $P$
 and odd under time reversal $T$, as follows from Eq.\ (\ref{basic}).
A pairwise exchange interaction hamiltonian, therefore,  is sufficient
for the
 present discussion. One can take ${\cal H} = {1\over{2}} \int d{\bf r}
 d{\bf r'} K({\bf r},{\bf r'})~{\bf S}({\bf r})\cdot {\bf S}({\bf r'}) $.
 Following the same approach as for the CSL case, we find for the time
 derivative $\partial_{t}B$
 \begin{eqnarray}
 \partial_{t}B({\bf r},{\bf r'}|0) = \int
 d{\bf r''} K({\bf r},{\bf r''})
 \{ \langle [{\bf S}({\bf r''})\cdot {\bf S}({\bf
 r'})] {\bf S}({\bf r})\rangle - {\bf S}^2 \langle{\bf S}({\bf
 r''})\rangle \}.
 \label{beq}
 \end{eqnarray}
Therefore, we conclude that a nonzero $B$-term in the decomposition
 of Eq.\ (\ref{decomposition}) describes a vector ordering $\langle{\bf
S}({\bf
 r})\rangle \neq 0 $. This is trivial ferromagnetic order. In particular,
for
 ${\bf r} = {\bf r'}$ it follows from Eq.\ (\ref{beq}) that
$~\partial_{t}B({\bf r},{\bf r}|0)
 \propto \langle{\bf S}({\bf
 r})\rangle$. The ferromagnet corresponds to the reducible three-spin
 correlator in Eq.\ (\ref{deriv}) and can be described in terms of a single
 spin operator expectation value. This contradicts our initial
restriction to states having no average spin density and we do not consider
this
ferromagnetic case any further.

3). {\em Chiral Spin Nematic}. As remarked earlier, the last term in the
decomposition of
Eq.\ (\ref{decomposition}) represents a spin
nematic order belonging to the tensor ($\l = 2$) representation of $O(3)$. The
crucial difference from previously considered spin nematics
\cite{AG} is that in our case $ Q_{ij}({\bf r},{\bf r'}|t)$ is an odd
function of $t$ and therefore it is an odd function under $P$.
In the  presence of an external
magnetic field, which expressly violates time reversal and spin rotation
symmetries, a $P$-odd nematic state was considered by
Chubukov \cite{chub}. In our considerations, however, the external magnetic
field is zero.

The parity
operation is
equivalent to interchange  of ${\bf r}, {\bf r'}$. The conventional spin
nematic with $ Q_{ij}({\bf r},{\bf r'})$ considered at equal times   is
even under $P$, as follows from Eq.\ (\ref{basic}) for $\Lambda_{ij}({\bf
r},{\bf r'}|t) = Q_{ij}({\bf r},{\bf r'}|t)$. We emphasize this distinction
because it shows that
the CSN state cannot be obtained within the old classification scheme \cite{AG}
and
is a new type of state.

As before, we take the time derivative to write the equal-time correlator. We
find
\begin{equation}
L_{ij}({\bf r},{\bf
r'}) \equiv \partial_{t}Q_{ij}({\bf r},{\bf r'}|0) =
\int d{\bf r''}K({\bf r},{\bf r''})
\epsilon_{iln} V_{lnj}({\bf r},{\bf r''},{\bf r'}).
\label{qeq}
\end{equation}
Here $V_{lnj}({\bf r},{\bf r''},{\bf r'}) = \langle S_{l}({\bf r})
S_{n}({\bf r''})
 S_{j}({\bf r'}) \rangle$ is the three-spin correlator and an isotropic
tensor $K_{ij} = \delta_{ij} K$ was taken. As mentioned above, only
the $P$-odd part of $K({\bf r},{\bf r''})$ contributes to the
integral and this constraint requires special models for exchange interactions.

The transformation properties of $V_{lnj}$ under permutation of the spin
indices
are given by a particular representation of the permutation group of three
objects ${\cal S}_3$ and in this particular case it is the $\l = 2$
representation. The symmetry of this tensor can be seen from
Eq.\ (\ref{qeq}): $V_{lnj} = V_{[ln]j}$, where the square brackets stand for
antisymmetrization with respect to the indices contained within. The
even-time totally symmetric three-spin correlator has been considered by
Gor'kov \cite{G}. This symmetry is obvious from the fact that the product
$\epsilon_{iln}V_{lnj}$ enters Eq.\ (\ref{qeq}). On the other hand, the time
derivative
$\partial_{t}Q_{ij}  = L_{ij}$ yields the symmetric traceless tensor $L_{ij}$.
Therefore,  the product $\epsilon_{iln}V_{lnj}$ should be symmetrized with
respect to
$i,j$. In what follows, we limit ourselves to the case of uniaxial solutions
for the CSN. An allowed solution is
\begin{eqnarray}
L_{ij}({\bf r},{\bf r'}) = L({\bf r},{\bf r'})(n_in_j -
{1\over{3}}\delta_{ij}),
\label{uniax}
\end{eqnarray}
where ${\bf n}$ is the nematic director (${\bf n}$ and $-{\bf n}$ are
equivalent)  in spin space. The director specifies
the plane of the spin modulation, as in the conventional spin nematic.
Although the form of $L_{ij}$ in the
above equation is identical to that for the conventional even-time spin nematic
with $Q_{ij}({\bf r},{\bf r'}) = Q({\bf r},{\bf r'}) (n_in_j -
{1\over{3}}\delta_{ij})$ \cite{AG}, here the physics is fundamentally
different. The odd-time $L_{ij}$ is inherently connected to
the three spin correlator  $V_{lnj}({\bf r},{\bf r''},{\bf r'})$ in
Eq.\ (\ref{qeq}), in contrast to a two spin correlator as in the conventional
spin nematic. In addition, the spatial parity of $L_{ij}$ is opposite to the
parity
 of the standard spin nematic  $Q_{ij}$.

In the exchange approximation, the coordinate and spin indices are
independent and one can factorize $V_{ljm} : \langle S_l({\bf r}_1)
S_n({\bf r}_2) S_j({\bf r}_3) \rangle  = T_{lnj} \Phi({\bf r}_1,{\bf
r}_2,{\bf r}_3)$.   The spin part is $T_{lnj} = \epsilon_{iln} (n_in_j -
{1\over{3}}\delta_{ij})$, and the orbital part $\Phi$ is to be integrated with
$K({\bf r},{\bf r'})$ in Eq.\ (\ref{qeq}) to yield $L({\bf r},{\bf r'})$ One
finds
\begin{eqnarray}
Q_{ij}({\bf r},{\bf r'}|t) &=& Q({\bf r},{\bf r'}|t)(n_in_j -
{1\over{3}}\delta_{ij}),\nonumber\\
\partial_{t}Q({\bf r},{\bf r'}|t) &=& L({\bf r},{\bf r'}).
\label{qsolution}
\end{eqnarray}
The lagrangian for the CSN state is identical to the lagrangian of  the
conventional spin nematics \cite{AG} and is
\begin{equation}
{\cal L} = {1 \over 2\gamma^2}\chi_{ij}({\bf n} \times \partial_t{\bf n}
+ \gamma {\bf H})_i({\bf n} \times \partial_t{\bf n} + \gamma {\bf H})_j -
J_{kl}
\partial_k{\bf n}\partial_l{\bf n} - U_{anis},
\end{equation}
where $\gamma$ is the gyromagnetic ratio, ${\bf H}$ is the external magnetic
field, $J_{ij}$ is the inhomogeneous exchange tensor, $\chi_{ij} = \chi_{\perp}
(\delta_{ij} + n_in_j)$ is the transverse susceptibilty tensor , and $U_{anis}$
is the relativistic anisotropy energy. Linear gradient terms are not allowed in
the lagrangian.

We mention briefly possible ways to generate a CSN phase. To
lower the symmetry of the system down to the CSN  one can
consider the quadrupolar interaction in the CSL phase ($\l  =
0$), which will generate the nematic ($\l  = 2$) component. In
the $P,T$ violating ground state this nematic state will have a
chiral component. Another possibility is spontaneous breakdown of
$P$ and $T$ in an already existing nematic state.

This work was supported in part by a J.R.
Oppenheimer fellowship (AVB) and by NSF Grant DMR92-21907(EA). Part of the
work presented here was performed at the Aspen Center for Physics, and
part at Los Alamos under the auspices of the Program on Strongly
Correlated Electron Systems, whose support is acknowledged.


\begin{references}
\bibitem{comment1}
Here the simplest possible model has been assumed to illustrate the
symmetry statements to be made about different types of odd-time magnetic
states. The microscopic origin of the model Hamiltonian (\ref{hamcsl})
will not be considered here.  More complicated Hamiltonians  are possible,
such as to include other fields, i.e. electrons, in the problem. Also, for
the higher spins another higher order invariants are possible as well.  We
will not address these questions here either.
\end{references}
\end{document}